\documentclass[english]{cccconf}
\usepackage[comma,numbers,square,sort&compress]{natbib}
\usepackage{amsmath}
\begin{document}

\title{A Framework of Stability Analysis for Multi-agent Systems on Any Topology Graph: Linear Systems}
%\author{WANG Yong \aref{amss},
%        LI Guiming Li\aref{amss,hit},
\author{WANG Yong, LI Guiming}

% Note: the first argument in the \affiliation command is optional.
% It defines a label for the affiliation which can be used in the \aref
% command. If there is only one affiliation for all authors, then the
% optional argument in the \affiliation command should be suppressed,
% and the \aref command should aslo be removed after each author in
% \author command, in this case the affiliation will not be numbered.

% 请注意：\affiliation命令的第一个参数是可选的，它定义了用于\aref命令的标签。
% 如果所有作者只有一个单位，请不要使用\affiliation命令的可选参数，同时在上面
% 的\author命令中的每位作者姓名后面也不能使用\aref命令，示例如下
% \author{First Author, Second Author, Third Author}
% \affiliation{Chinese Academy of Sciences, Beijing 100190, P.~R.~China\email{ccc@amss.ac.cn}}
% 此时单位前不会有数字编号，作者姓名后面也没有编号

\affiliation{Beijing Institute of Control Engineering,
         Beijing 100190, P.~R.~China\email{wyongzju@163.com}}

%\affiliation[amss]{Beijing Institute of Control Engineering,
%         Beijing 100190, P.~R.~China
%        \email{ccc@amss.ac.cn}}
%\affiliation[hit]{Harbin Institute of Technology, Harbin 150001, P.~R.~China
%        \email{xxx@hit.edu.cn}}

\maketitle

\begin{abstract}
In this paper, from the structural perspective, we propose a new stability analysis approach for the consensus of linear multi-agent systems. Different from the general tools: the Laplacian matrix based method and the Lyapunov's method, this  approach  treats the multi-agent system as the composition of many isolated agents, and focuses on their special  input and output relationship.  Through transforming the construction of a graph into a standard procedure only including three basic structures, the stability analysis is recursive and   independent of the specific topology. Therefore, this approach can be used for multi-agent systems on  any topology graph.
\end{abstract}


% Please remove or comment out the following line if the footnote is not necessary
\footnotetext{This research is supported by the National Natural Science Foundation of China under grant 61333008 and the National Basic Research
Program (973) of China under Grant 2013CB733100.}

\section{Introduction}
In this decade, the problem of coordination and control of multi-agent systems have attracted more and more attentions.  Researchers study a variety of interesting problems from different angles. At first, the research subjects only involve the single-integrator and the double-integrator linear system,e.g.,\cite{Ren2008} and \cite{Mesbahi2010}, and then extend to the high-order linear and nonlinear system,e.g.,[3],\cite{Kim2011},and \cite{Xi2012}. The model of multi-agent systems evolves from the continuous-time system to the discrete-time system,e.g.,\cite{Su2012}. And the topology of multi-agent systems is investigated from the simple fixed graph to the switching time-varying graph,e.g.,\cite{Ren2005} and \cite{Olfati2004}. However, there are only two methods to be widely used to analyze the consensus for  almost all of above systems, i.e., the Laplacian matrix  based method and the Lyapunov function based method. The first method belongs to the algebraic graph theory, which introduces the Laplacian matrix $L_p$ to the multi-agent system. Since the Laplacian matrix $L_p$ is better able to reflect the topology information of the multi-agent system, and its character that it has a simple zero eigenvalue with an associated eigenvector $1_p$  and all other eigenvalues have positive real parts\cite{Ren2008}, uncovers the essence of the multi-agent system being able to achieve consensus, it is widely used in the linear multi-agent system.  The seconde method is widely used in the nonlinear system, but it still faces the difficulty how to  find a  Lyapunov function candidate for the general system. Furthermore,  the complex topology of multi-agent systems causes  its extra difficulty. Therefore, the successful application of the Lyapunov function based method is confined to some special topology graphs, e.g., the undirected, balanced or strongly connected graphs in \cite{Chopra2012},\cite{Listmann2009},\cite{Das2010},\cite{Nosrati2012},and \cite{Yu2011}.

It should be mentioned that above two methods have a common character, that is both of them treat the multi-agent system as a whole and consider its all details. In this sense, the complex topology of  multi-agent systems make it more hard to find a unified approach to design the Lyapunov function candidate for  any tropology graph. However, if we change the viewpoint and look at the problem from a structural perspective, a different approach can also be used to solve this problem. Based on the fact that every agent in the multi-agent system is ISS, in this paper we will propose a new consensus analysis approach for the multi-agent system on any topology graph. Compare with the old framework, this approach treats the multi-agent system as the composition of many isolated agents, ignores their internal details, and focuses on their special  input and output relationship. In this way, the analysis is a recursive and constructive process. To sum up, this paper has three contributions. Firstly, we give a new consensus analysis approach for the multi-agent system. It notes that this method bases on the fact that every agent is ISS and is different from the methods in the current literatures.  Secondly, this method is unified for linear and nonlinear  multi-agent systems on any topology graph. In contrast, the  Laplacian matrix  based method is suitable for any topology graph, but is just used in linear systems. The Lyapunov function based method can be used for linear or nonlinear systems, but is too hard to find a unified Lyapunov function candidate for any topology multi-agent system. Thirdly,  this approach uncovers the essence of how multi-agent systems achieving consensus from another perspective. In the present literatures, for linear systems, the consensus is able to be achieved  due to the  special character of  engenvalues of $L_p$,  and for nonlinear systems, the zero points of  Lyapunov function are a continuous trajectory . In this paper, we will give a structural explanation, that any topology graph with a spanning tree can be transformed into a cascade structure where the first agent converges to a constant determined by the initial value of  some agents. In order easily to  illustrate the main framework, this paper only investigates  the single-integrator linear system, and  more complex multi-agent systems, e.g.,the nonlinear system, will be studied later on. Then the rest of paper is organized as follows. Section 2 briefly recalls some basic background knowledge about the graph and consensus problem in the multi-agent system. Section 3 gives an approach to construct a graph containing a spanning tree. Section 4 presents three basic structures and studies their consensus properties. Section 5 analyzes the consensus of the multi-agent system. Section 6 is the conclusion.

\section{BACKGROUND}

\subsection{Graph Theory }

Firstly, some graph terminologies and notions are introduced which can be seen in \cite{Mesbahi2010}.
A graph is a pair $G({V_p},{E_p})$,where ${V_p} = \{ {v_1},...,{v_p}\} ,p \in N$ is a finite nonempty node set ,$\{ {v_i},{v_j}\} $denotes an edge and $E_p$ denotes the edge  set of ordered pairs of nodes, called edges. If the ordered pair $({v_i},{v_j}) \in {E_p}$, then $v_i$ is said to be the head (where the arrow starts) of the edge, while $v_j$ is its tail. If, for all $({v_i},{v_j}) \in {E_p}$, we have $({v_j},{v_i}) \in {E_p}$, then the graph is said to be undirected. Otherwise, it is called the directed graph. An edge $({v_i},{v_j})$ is said to be incoming with respect to $v_j$  and outgoing with respect to $v_i$  and can be represented as an arrow with vertex $v_i$  as its tail and vertex $v_j$  as its head. A path of length $r$ in a directed graph is a sequence  ${v_0},...,{v_r}$ of $r+1$ distinct vertices such that for every $i \in \{ 0,...,r - 1\} $ ,$({v_i},{v_{i + 1}})$  is an edge. A directed graph is strongly connected if any two vertices can be joined by a path. A graph is balanced, if for each ${v_i} \in V$ , the in-degree equals the out-degree. Clearly every undirected graph is balanced.

Define the Laplacian matrix  ${L_p} = [{l_{ij}}] \in {{\bf{R}}^{p \times p}}$  as

${l_{ii}} = \sum\limits_{j = 1,j \ne i}^n {{a_{ij}}} $ and ${l_{ij}} =  - {a_{ij}},i \ne j$

Note that if $({v_i},{v_j}) \notin {E_p}$ then ${l_{ij}} =  - {a_{ij}} = 0$ .Matrix $L_p$  satisfies
${l_{ij}} < 0,i \ne j,\sum\limits_{j = 1}^n {{l_{ij}}}  = 0,i = 1,...,p.$

\subsection{Consensus Problem }

Consider the general multi-agent system with single-integrator dynamics given by
\begin{equation}  \label{1}
{\dot x_i} = u_i,i = 1,...,n.
\end{equation}
The consensus problem of (1) can be seen as a graph problem,  and the classical continuous-time consensus algorithm is
\begin{equation}  \label{2}
u_i = \sum\limits_{j \in {N_i}}^{} {{a_{ji}}({x_j} - {x_i})}.
\end{equation}
where ${a_{ji}} > 0$ .The controller can be described with the Laplacian matrix, and the  compact form is
\begin{equation}  \label{3}
\dot x =  - {L_n}x.
\end{equation}
We say that consensus is achieved when
\[\mathop {\lim }\limits_{t \to \infty } \left\| {{x_i}(t) - {x_j}(t)} \right\| = 0.\]
Thus, the properties of the Laplacian matrix  determine the behavior of  (1).

Next, we will propose a new approach to analyze the con-sensus problem of system (1) with the protocol (2).

\section{APPROACH TO CONSTRUCT A GRAPH CONTAINING A SPANNING TREE}
In this section, we will present a standard procedure to reconstruct any topology graph with a spanning tree by the following recursive process:

{\textbf{Recursive process 1.}

Adding one vertex  to the (k-1)-th graph ${{\bf{G}}^{k - 1}}$ in one
of  3 basic topologies yields the k-th graph  ${{\bf{G}}^{k }}$.

where the initial graph is a vertex, and the three basic topologies are the cascade structure, the interconnected structure and the blended structure of  them given by  Fig.1.

\begin{center}
\includegraphics [scale=0.3]{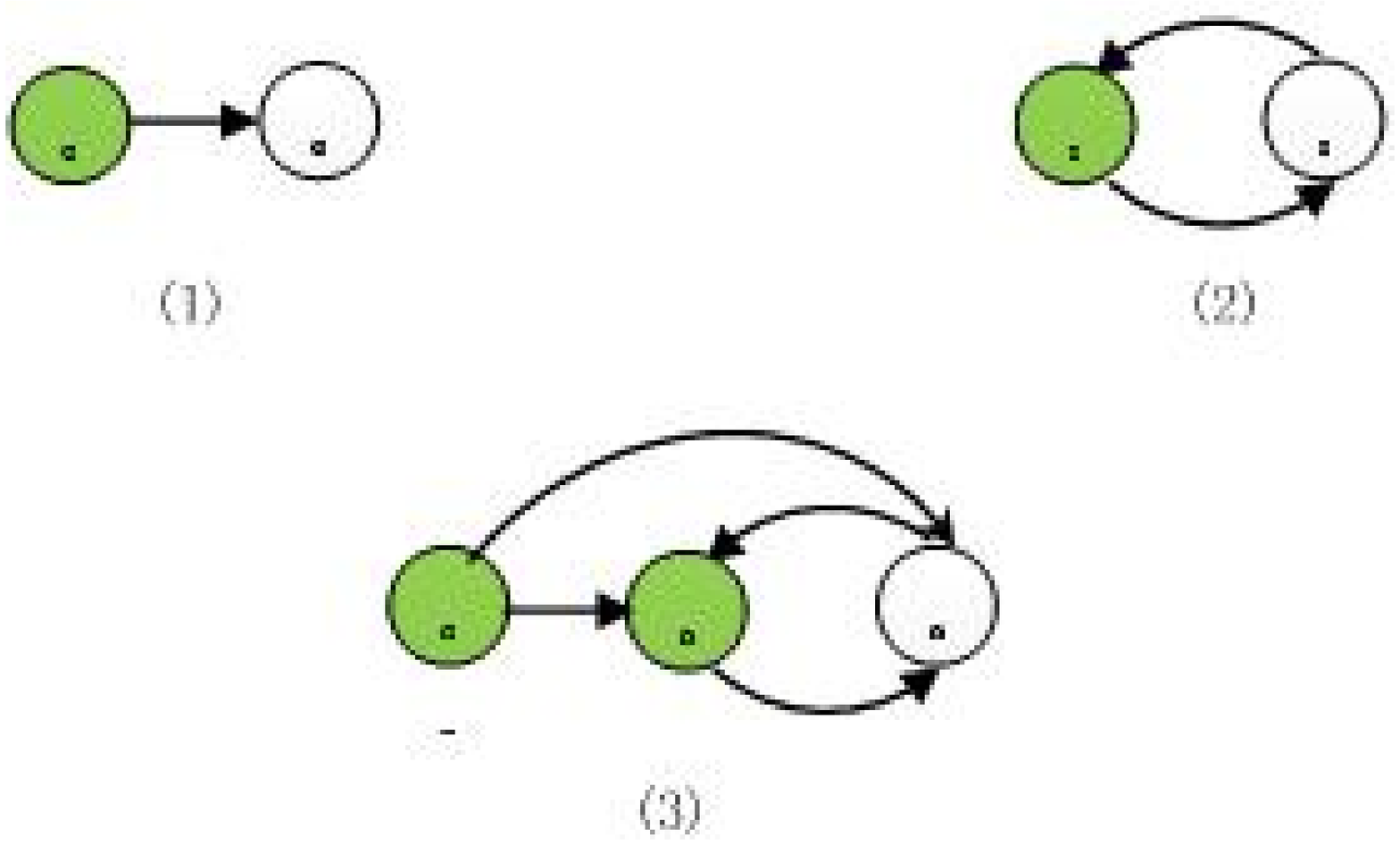}
\\
{\fontsize{7.3pt}{11.6pt}\selectfont
Fig.~1~~(1) is the cascade structure, (2) is the interconnected structure, and (3) is the blended structure, where the shadowed circle denotes the graph, and the blank denotes the vertex. }
\end{center}

%\begin{figure}
%\includegraphics[scale=0.3 ]{fig1.eps}
%\caption{(1) is the cascade structure, (2) is the interconnected structure, and (3) is the blended structure, where the shadowed circle denotes the graph, and the blank denotes the vertex.}\label{fig:side:a}
%\end{figure}

%\begin{center}
%\includegraphics [scale=0.6]{fig1.jpg}
%{\fontsize{7.3pt}{11.6pt}\selectfont Fig.~4~~(1) is the cascade structure, (2) is the interconnected structure, and (3) is the blended structure, where the sha-dowed circle denotes the graph, and the blank denotes the vertex. }
%\end{center}

Before presenting  the procedure, we first give some notions.

${\bf{g}} = (V,E)$ denotes a graph with $n$  vertices, which contains a spanning tree denoted as ${{\bf{g}}_s} = ({V_s},{E_s})$ . Since ${V_s} = V$ and ${E_s} \subseteq E$ , define the set of edges ${E_{\bar s}} = \{ e \in E|e \notin {E_s}\} $ such that  ${E_{\bar s}} \cup {E_s} = E$ and ${E_{\bar s}} \cap {E_s} = \emptyset $ .  ${{\bf{G}}^k} = ({V^k},{E^k})$ denotes the new graph in the $k{\rm{ - }}th$ step of the procedure, and   ${\bf{G}}_s^k = (V_s^k,E_s^k)$ denotes the associated growing-up spanning tree in the $k{\rm{ - }}th$ step. $E_i^{head} = \{ ({v_i},{v_j}) \in E|{v_j} \in G\} $ denotes a set of order edges where the common head of the edges is the vertex  $v_i$ and their tails belong to a group $G$ ,otherwise, $E_j^{tail} = \{ ({v_i},{v_j}) \in E|{v_i} \in G\} $ denotes a set of order edges where the common tail  of the edges is the vertex $v_i$ and  their heads belong to a group $G$ .For $V_s$ , we renumber every vertex consecutively according to the width-first sequence,  e.g.,  $v_j^i$ denotes the  $j$-th child from left to right of  $i$-th level from top to bottom. Then following this rule, the root can be numbered as $v_1^1$  ,the first child of root node is numbered  as  $v_2^1$   ,and  the second is $v_2^2$   , and so on. In this way, the entire graph with a spanning tree can be reconstructed by the following procedure.

\textbf{Procedure 1.}

\textbf{Initialization.} Arbitrarily choose one of spanning trees contained in the graph ${\bf{g}} = (V,E)$   and denote it as ${{\bf{g}}_s} = ({V_s},{E_s})$ .  Set $E_{\bar s}^0 = E - {E_s}$ , ${{\bf{G}}^0} = \emptyset $ , and  ${\bf{G}}_s^0 = \emptyset $.

\textbf{Step 1.}  Let the root $v_1^1$  of $g_s$ as the first element of $G^1$  and  $G_1^s$ such that  $
{\bf{G}}_s^1 = {{\bf{G}}^1}$.

\textbf{Step 2. }  Choosing the  $k$-th vertex $v_i^j$  in $g_s$  according to the width-first sequence , and  adding it to $G^{k-1}$  and $G_s^{k-1}$  yields $G^{k}$   and $G_s^{k}$ , where $k = 2,...,n$ . Then choosing order edges $E_j^{head} = \{ (v_j^i,v_p^q) \in E|v_p^q \in {{\bf{G}}^{k - 1}}\} $ and adding them to  $G^{k-1}$  yields ${{\bf{\bar G}}^k}$ .

\textbf{Step 3.} Choosing order edges $E_j^{tail} = \{ (v_p^q,v_j^i) \in E|v_p^q \in {{\bf{G}}^{k - 1}}\} $ and adding them to ${{\bf{\bar G}}^k}$ yields the set of edges ${E_j} = E_j^{head} \cup E_j^{tail}$, and  $\{ (v_i^j,v_p^q) \in E_{\bar s}^k|v_p^q \in {{\bf{G}}^{k - 1}}\}  \subseteq {E_j}$. Remove them from $E_{\bar s}^{k - 1}$ so as to obtain  a new graph  $G^k$ and a new edge set $E_{\bar s}^{k }$ .

\textbf{Step 4.}  Repeat step 2 and 3  till $E_{\bar s}^n = \emptyset $ so that a new graph  ${{\bf{G}}^n} = ({V^n},{E^n})$ is reconstructed.

\textbf{Example 1.} By the above procedure, a  graph    including five vertices can be reconstructed as follows.

\begin{center}
\includegraphics [scale=0.15]{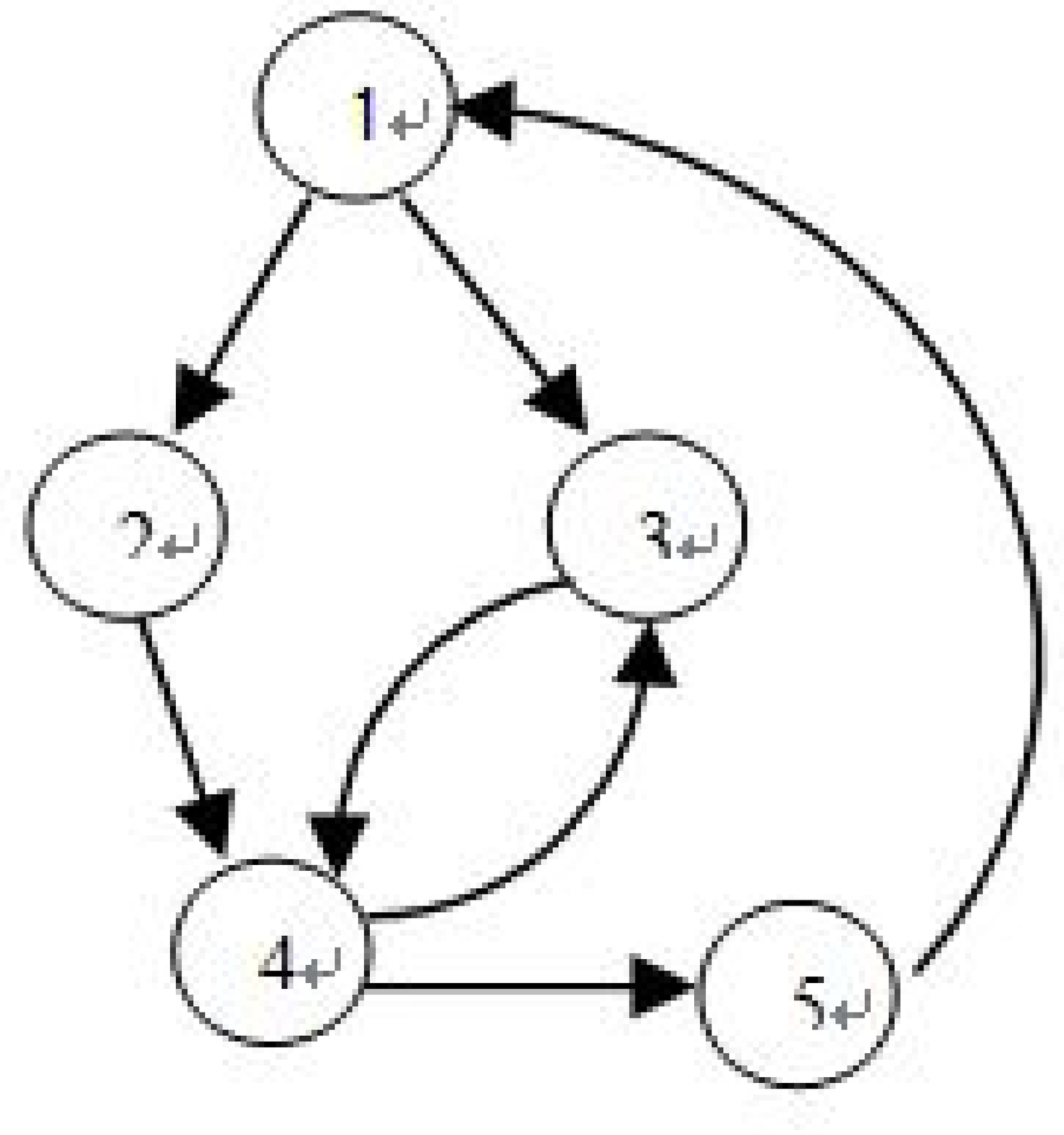}
\includegraphics [scale=0.15]{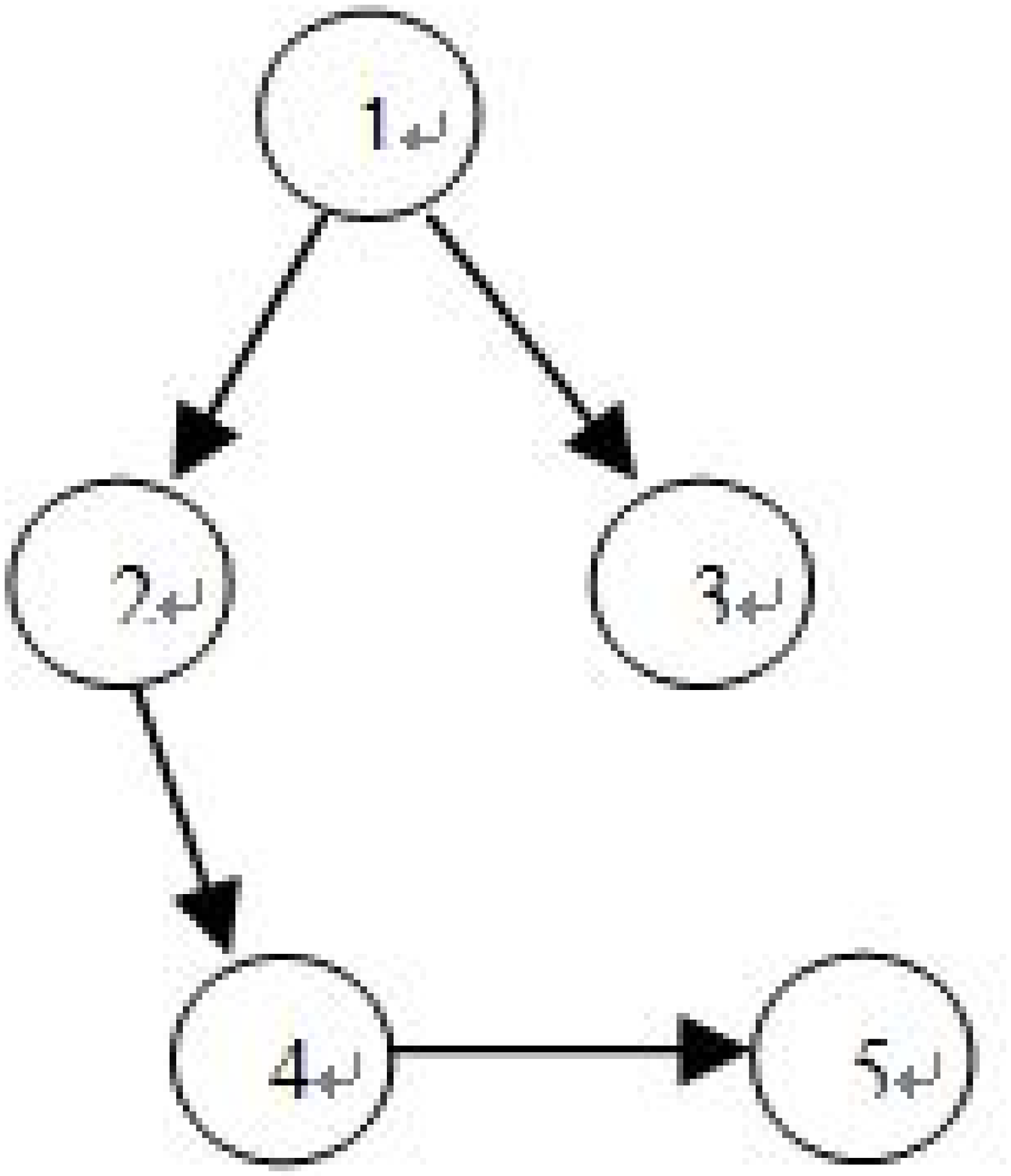}
\\
{\fontsize{7.3pt}{11.6pt}\selectfont
Fig.~2~~a graph $g=(V,E)$  ~~~~~~               Fig.~3~~one spanning tree $g_s=(V_s,E_s)$  }
\end{center}

\begin{center}
\includegraphics [scale=0.2]{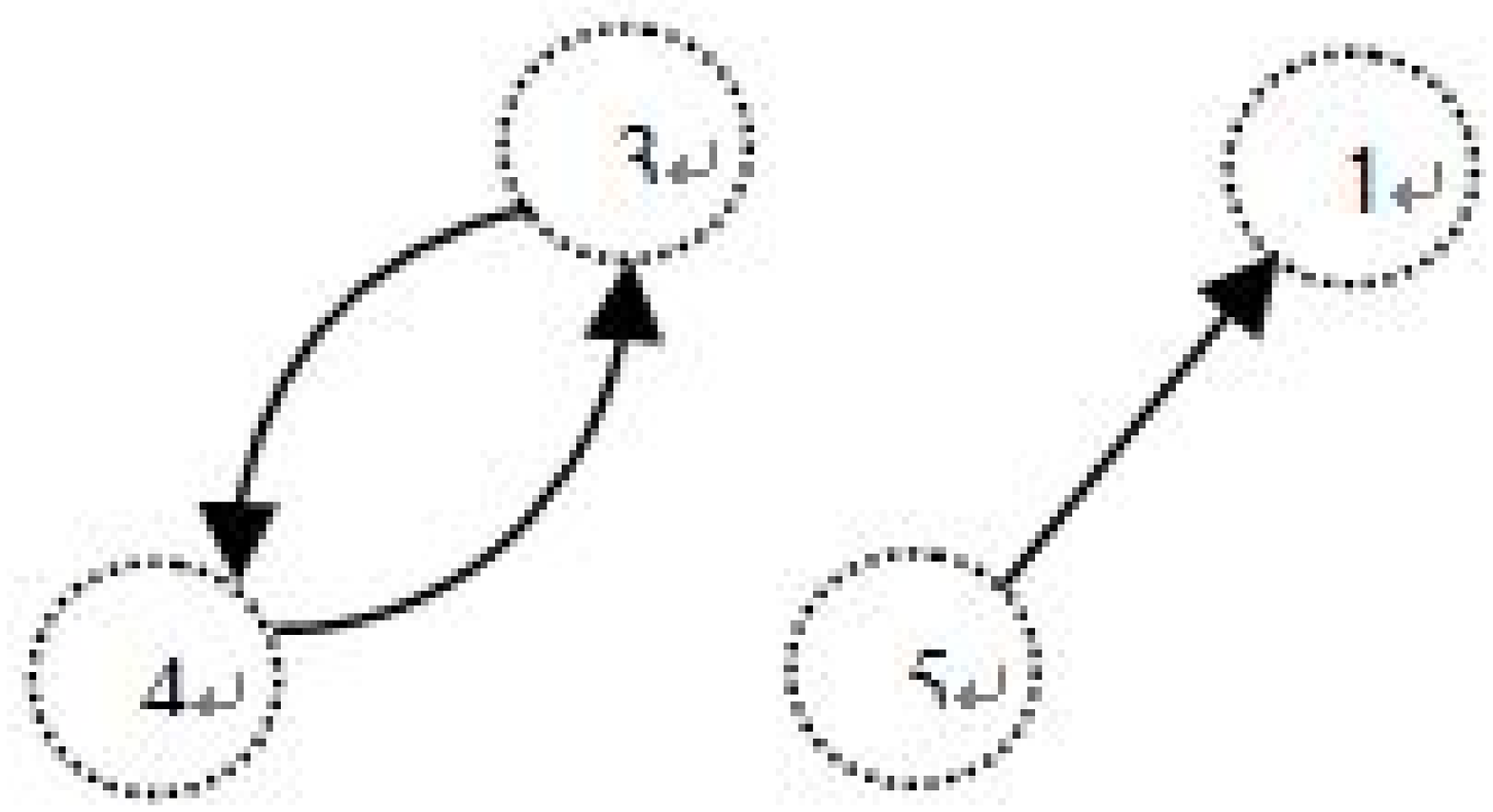}
\\
{\fontsize{7.3pt}{11.6pt}\selectfont
Fig.~4~~$E_{\bar s}^0$ }
\end{center}

\begin{center}
\includegraphics [scale=0.3]{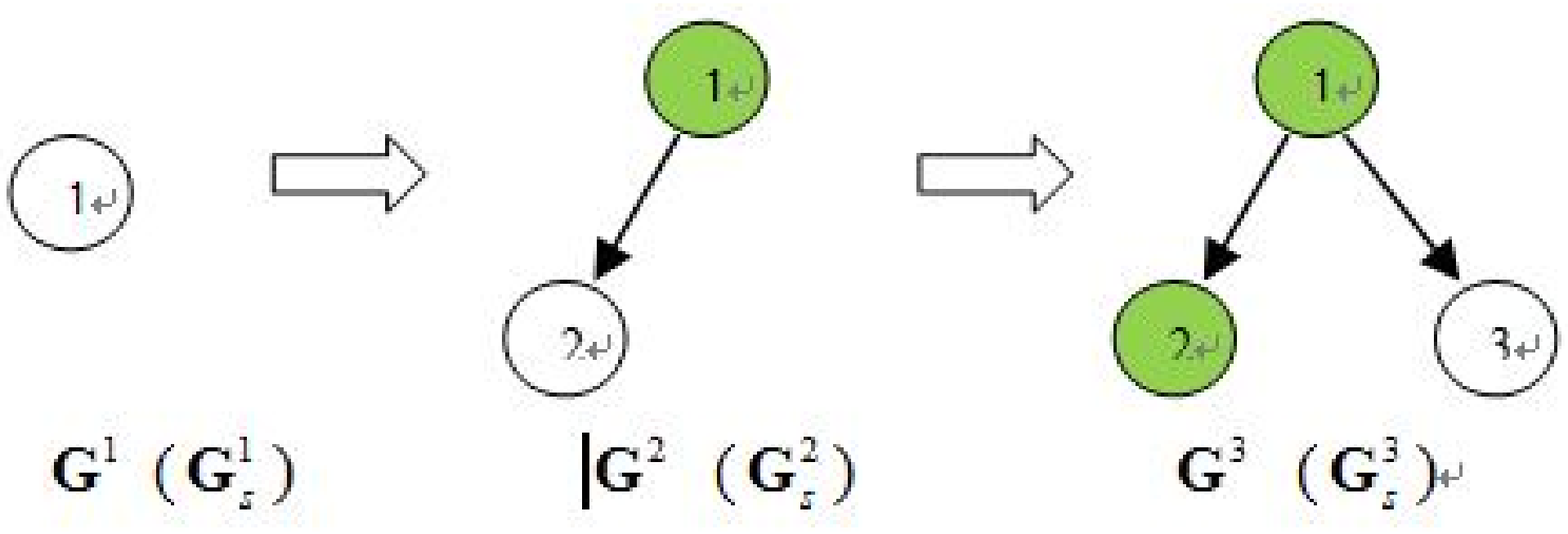}
\\
\includegraphics [scale=0.3]{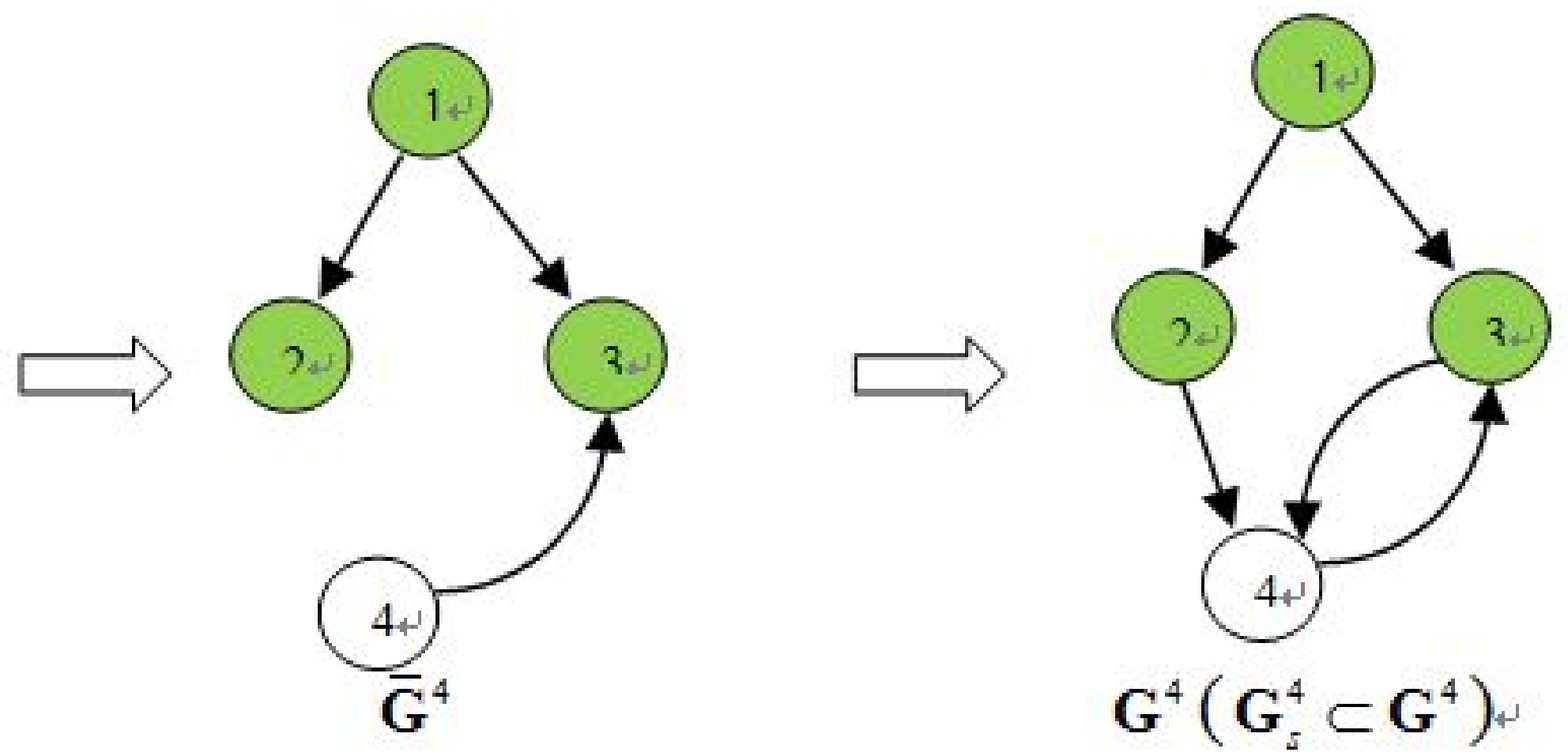}
\\
\includegraphics [scale=0.3]{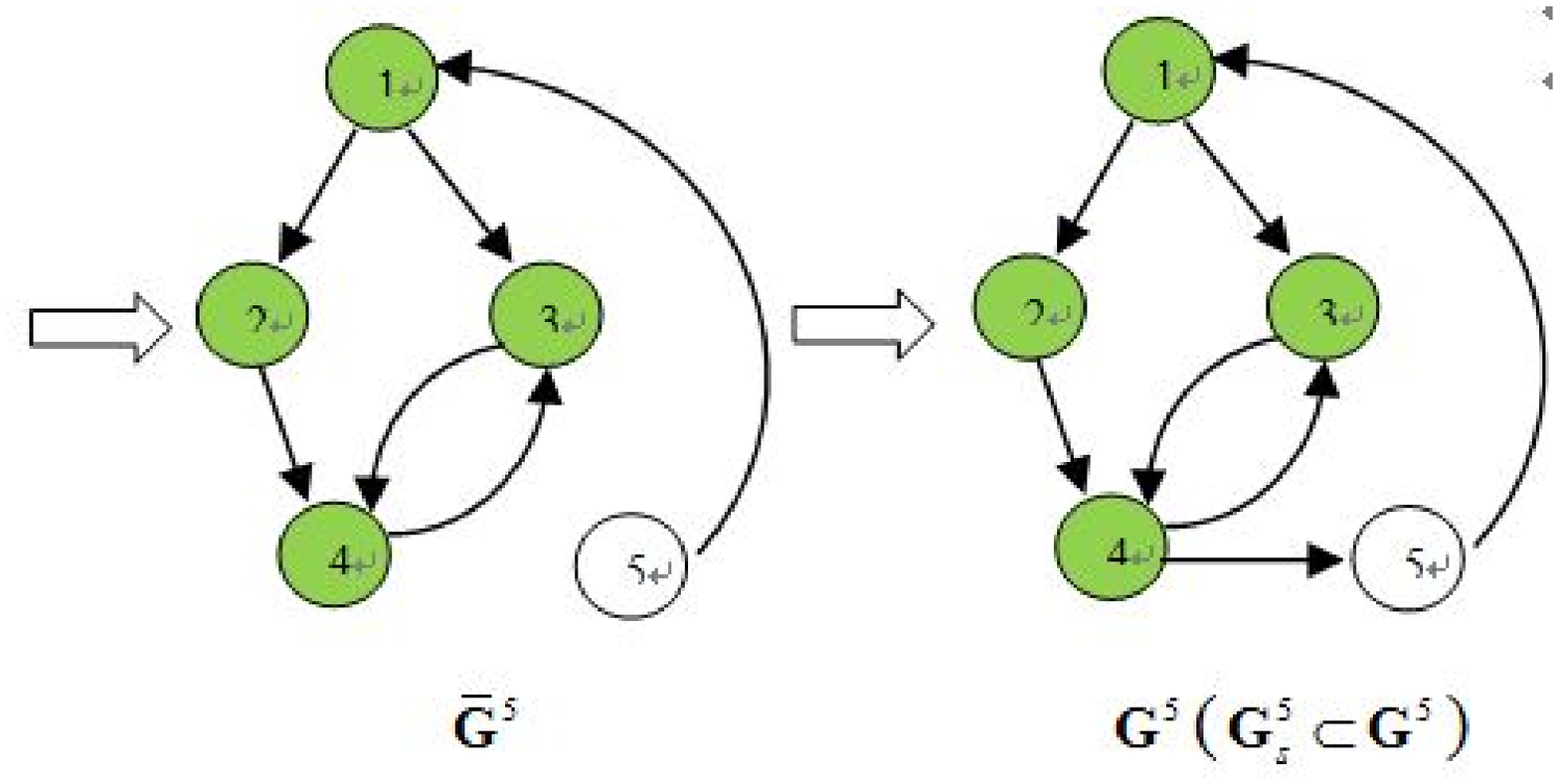}
\\
{\fontsize{7.3pt}{11.6pt}\selectfont
Fig.~5~~Add $v^i_j$  to $G^{k-1}$ and $G_s^{k-1}$ , then obtain  $G^{k}$ and $G_s^{k}$ . }
\end{center}

%
%\begin{figure}
%\includegraphics[scale=0.3 ]{fig5.1.eps}
%\caption{.}\label{fig:side:a}
%\end{figure}
%%
%\begin{figure}
%\includegraphics[scale=0.3 ]{fig5.2.eps}
%\caption{}\label{fig:side:a}
%\end{figure}
%
%\begin{figure}
%\includegraphics[scale=0.3 ]{fig5.3.eps}
%\caption{ add $v^i_j$  to $G^{k-1}$ and $G_s^{k-1}$ , then obtain  $G^{k}$ and $G_s^{k}$ . $G^{2}$ and $G^{3}$ are the cascade structure, and $G^{4}$  is a blended structure. $G^{5}$ is an interconnected structure.}\label{fig:side:a}
%\end{figure}
%
%
%

\emph{\textbf{Remark 1.}}
(1) The graph $g$  in the Fig.2 has many spanning trees which contains all vertices and part of edges of $g$. $g_s$  in Fig.3 is just one of them.

(2)  In the Fig.4, $E_{\bar s}^0$ is the rest edges after transforming $g$  into $g_s$ ,which will be added to $G^k$  later.

(3) In the Fig.5, at every step, $G^1$ and $v_2^1$ (vertex 2),  $G^2$ and $v_2^2$  (vertex 3) make up the cascade structure;  $G^3$ and $v_3^1$   (vertex 4) make up the blended structure;  $G^4$ and $v_4^1$  (vertex 5) make up the interconnected structure.

(4) At every step,  $G^k$ always has a spanning tree, which guarantees the multi-agent system achieves consensus.

The following proposition aims to  show any topology graph containing a spanning tree can be reconstructed by the above  procedure .

\textbf{Proposition 1.}  Any graph containing a spanning tree can be reconstructed by the procedure 1, and the new graph in every step contains a spanning tree.

\emph{Proof. }At first, suppose  ${{\bf{g}}_s} = ({V_s},{E_s})$is one of  spanning trees in the graph $g=(V,E)$. Secondly, there are three cases to add a vertex $v_j^i$  to ${\bf{G}}_s^{k - 1}$ and ${\bf{G}}^{k - 1}$ along $g_s$  ,where  $k = 2,...n$, according to the width-first sequence :

\begin{description}
  \item[Case 1:]~~$G^{k-1}$ and $v^i_j$ make up the cascade structure , i.e., there are some directed edges from $G^{k-1}$   to  $v^i_j$.
  \item[Case 2:]~~$G^{k-1}$ and $v^i_j$  make up the interconnected structure. Based on the case 1, there are some directed edges from  $v^i_j$ to $G^{k-1}$  and  $v^i_j$  is the root of spanning trees in $G^{k}$.
  \item[Case 3:]~~$G^{k-1}$ and $v^i_j$   make up the blended structure . The configuration is similar with the case 2, but   $v^i_j$ is not the root of  any spanning tree in $G^{k}$.
\end{description}
 Finally, after adding the last vertex  and edges to $G^{n-1}_s$ and $G^{n-1}$  ,we have ${V^n} = {V_s} = V$   and ${E^n} = {E_{\bar s}} \cup {E_s} = E$ . It follows that ${\bf{G}}_s^n = {{\bf{g}}_s}$ and ${\bf{G}}_{}^n = {\bf{g}}$. Then the original graph has been reconstructed, and the above mechanism  guarantees that  $G^{k}$ in every step contains a spanning tree. %\blacksquare

\section{CONSENSUS OF BASIC TOPOLOGIES}

Following the above procedure, the consensus analysis is very simple. As long as we can prove the following recursive process, the consensus analysis of  multi-agent systems on any topology graph can be achieved.

{\textbf{Recursive process 2.}

{Adding} one agent to the consensus MAS $G^{k-1}$ {in} one of 3 basic topologies  {yields} a consensus  MAS $G^{k}$

In this section, we will introduce three basic topologies and prove the above recursive process.

\subsection{Input-State Pair}
Firstly, we will introduce the following notion.

{\textbf{Definition  1.}}Consider the scalar system
\begin{equation}  \label{13}
\dot v = \varphi (v,r)
\end{equation}
where $v,r \in {\bf{R}}$ , $\varphi :{\bf{R}} \times {\bf{R}} \to {\bf{R}}$ and meets the following properties:
\begin{description}
  \item[1)] $\varphi$ is differential mapping.
  \item[2)] $\varphi (v,r) = 0 \Leftrightarrow v = r$
  \item[3)] $\varphi (v,r) =  - \varphi (r,v)$
  \item[4)] $(v - r)\varphi (v,r) < 0,\forall v \ne r$
\end{description}
Then, the  pair $(v,r)$  is called Input-state pair (for short, ISP).

\emph{\textbf{Remark 2. }} This definition is very similar with the nonlinear protocol in the assumption 1 of \cite{Liu2009}. The  property 2) means the system (4) has an equilibrium point $r$ , which also implies the leader is an equilibrium point .  3) and 4) guarantee the stability of system (4). Essentially, ISP is a fundamental element of consensus protocols and the system with ISP is a simplest consensus system, where the leader is the input $r$  and the follower is the state variable $v$ .  More complex consensus protocols can be composed by many  input-state pairs ( for short,ISPS).

\emph{\textbf{Example 2.}}  ISP can form many consensus protocols, including the classical Laplacian matrices for linear system, and the nonlinear protocol in \cite{Das2010},\cite{Nosrati2012},\cite{Yu2011},and \cite{Liu2009}. Besides, some specific examples are  $- Kv + Kr$,$ - {v^3} + {r^3}$ , $ - \tan v + \tan r$, and $\tanh (v - r)$ .

\subsection{Cascade Structure}
It is well known that there are two basic kinds of structure for the complex system. One is the cascade structure and the other is the interconnected structure. This subsection will investigate the cascade structure in the multi-agent system background.

\begin{center}
\includegraphics [scale=0.3]{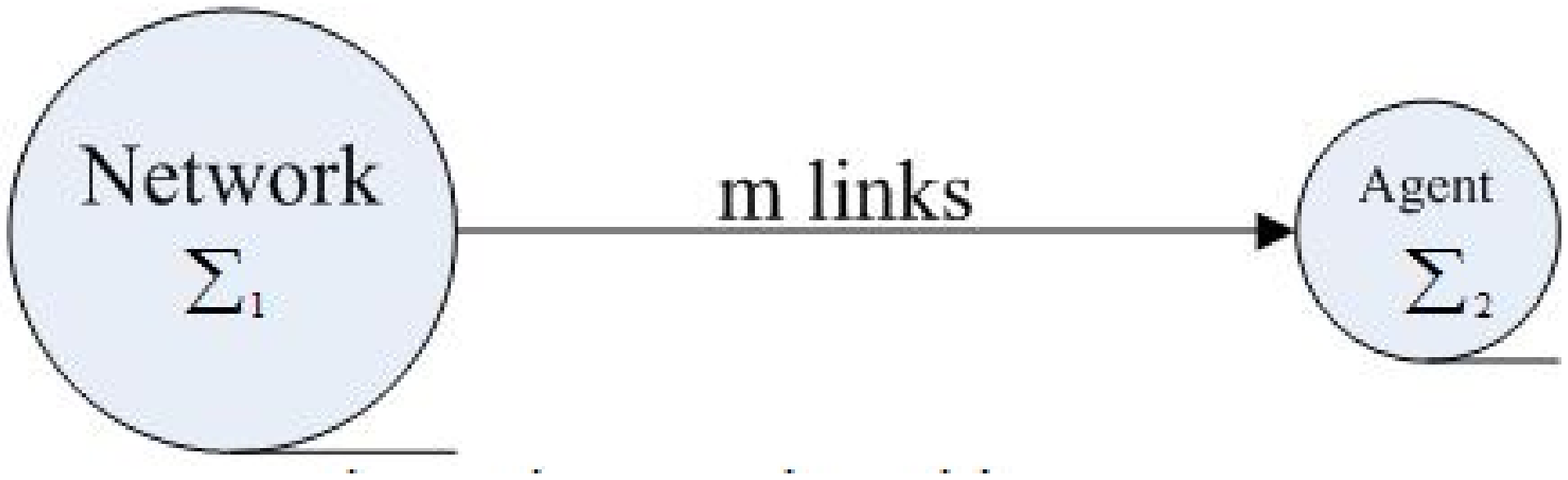}
\\
{\fontsize{7.3pt}{11.6pt}\selectfont
Fig.~6~~The cascade multi-agent system. }
\end{center}

%\begin{figure}
%\includegraphics[scale=0.3 ]{fig6.eps}
%\caption{ Fig.~6~~The cascade multi-agent system.}\label{fig:side:a}
%\end{figure}

Consider the multi-agent system $\sum_1 $ with $n$  agents where the agent dynamics are described by
\begin{equation}  \label{5}
{\dot x_i} = \sum\limits_{j \in {N_i}} {K_{ji}^{}({x_j} - {x_i})}
\end{equation}
There are  $m$ order edges from $\sum_1 $ to a scalar agent $\sum_2 $ modeled as
\begin{equation}  \label{6}
\dot z = \sum\limits_{p \in {N_z}}^{} {{K_{pz}}({x_p} - z)}
\end{equation}
where  ${x_i},{x_j},{x_p},z \in {\bf{R}}$, $K_{ji}^{},{K_{pz}} > 0$ , $i = 1,2,...,n$ ,$p = 1,2,...,m$  ,$m \le n$ . Suppose system $\sum_1 $ satisfies the following assumption.

{\textbf{Assumption 1.}} System $\Sigma_1$  can achieve consensus and converges to a constant $x_e$ which is determined by the initial value of $\Sigma_1$.

It follows that $\sum_1$  and $\sum_2$   make up a cascade structure. Corresponding to the recursive process 2, we have the following result.

{\textbf{Lemma 1.}} If the system $\sum_1$  satisfies the assumption 1 , the cascade system composed of $\sum_1$ and $\sum_2$ can achieve consensus.

Proof.  Solving equation (\ref{6}) yields
\begin{equation}  \label{7}
z(t) = \beta ({z_0},t) + \sum\limits_{p \in {N_z}}^{} {{K_{pz}}\int_0^t {{e^{ - K(t - s)}}{x_p}(s)ds} }
\end{equation}
where  $\beta ({z_0},t) = {e^{ - Kt}}{z_0}$ , $K = \sum\limits_{p \in {N_z}}^{} {{K_{pz}}} $.

By the assumption 1,  $\sum_1$ converges to a common value $x_e$ ,that implies there exists ${\beta _p} \in KL$  such that
\[\left\| {{x_p}(t) - {x_e}} \right\| \le {\beta _p}({x_0},t)\]
Equivalently, we have
\[ - {\beta _p}({x_0},t) + {x_e} \le {x_p}(t) \le {\beta _p}({x_0},t) + {x_e}\]
It follows that
\begin{align}\label{8}
&\sum\limits_{p \in {N_z}}^{} {{K_{pz}}\int_0^t {{e^{ - K(t - s)}}{x_p}(s)ds} } \nonumber \\
&\le K\int_0^t {{e^{ - K(t - s)}}{\beta _p}({x_0},s)ds}  + {x_e}K\int_0^t {{e^{ - K(t - s)}}ds} \nonumber \\
&= K\int_0^t {{e^{ - K(t - s)}}{\beta _p}({x_0},s)ds}- {e^{ - Kt}}{x_e} + x_e
\end{align}

and

\begin{align} \label{9}
&\sum\limits_{p \in {N_z}}^{} {{K_{pz}}\int_0^t {{e^{ - K(t - s)}}{x_p}(s)ds} } \nonumber \\
 &\ge  - K\int_0^t {{e^{ - K(t - s)}}{\beta _p}({x_0},s)ds}  + {e^{ - Kt}}{x_e}+ x_e
\end{align}

Substituting (\ref{9}) into (\ref{7}) yields
\[z(t) \le \beta ({z_0},t) + K\int_0^t {{e^{ - K(t - s)}}{\beta _p}({x_0},s)ds} - {e^{ - Kt}}{x_e} + {x_e}\]
\[z(t) \ge  - \beta ({z_0},t) - K\int_0^t {{e^{ - K(t - s)}}{\beta _p}({x_0},s)ds} - {e^{ - Kt}}{x_e} + {x_e}\]
Therefore, we have
\[\mathop {\lim }\limits_{t \to \infty } \left\| {z(t) - {x_e}} \right\| = 0\]
which implies
\[\mathop {\lim }\limits_{t \to \infty } \left\| {z(t) - {x_i}(t)} \right\| = 0\]
%\blacksquare

\subsection{Interconnected Structure}
In this subsection, we will talk about another structure , the interconnected structure, in the multi-agent system background.

\begin{center}
\includegraphics [scale=0.3]{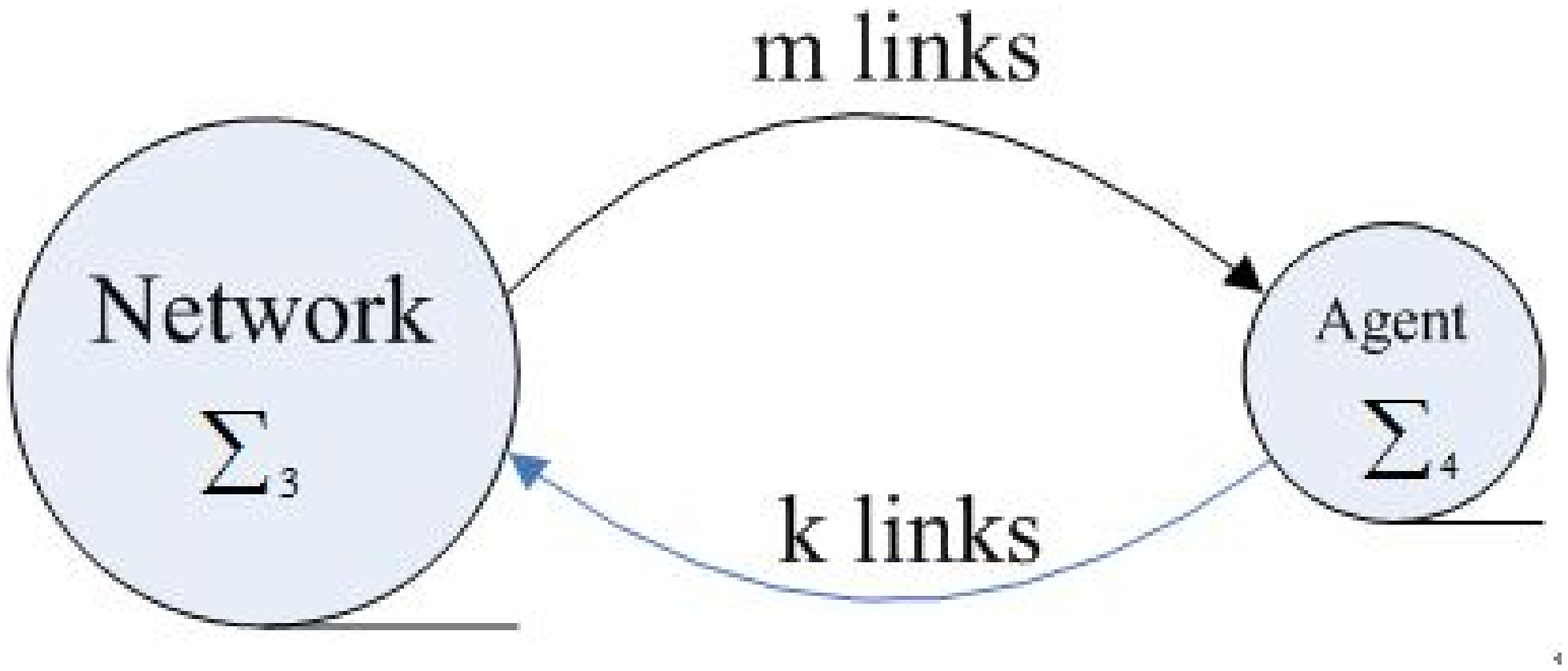}
\\
{\fontsize{7.3pt}{11.6pt}\selectfont
Fig.~7~~The interconnected multi-agent system. }
\end{center}

On the basis of the cascade structure, adding some order edges from $\sum_1$ to $\sum_2$ yields the interconnected system which can be seen as Fig.7. The associated new dynamics of $\sum_1$  and  $\sum_2$  can be respectively  written as follows.

$\sum_3$:
\[\left\{ \begin{array}{l}
{{\dot x}_1} = \sum\limits_{j \in {N_1}} {K_{j1}^{}({x_j} - {x_1})}  + K_{z1}(z - {x_1})\\
...\\
{{\dot x}_i} = \sum\limits_{j \in {N_i}} {K_{ji}^{}({x_j} - {x_i})}  + K_{zi}(z - {x_i})\\
{{\dot x}_{i + 1}} = \sum\limits_{j \in {N_{i + 1}}} {K_{j,i+1}^{}({x_j} - {x_{i + 1}})} \\
...\\
{{\dot x}_n} = \sum\limits_{j \in {N_{n
}}} {K_{jn}^{}({x_j} - {x_n})}
\end{array} \right. = Ax + Bz\]
$\sum_4$:
\[\dot z = \sum\limits_{q \in {N_z}}^{} {{K_{qz}}({x_q} - z)} \]

{\textbf{Assumption 2.}} There exists a spanning tree in the interconnected multi-agent system whose root is  $\sum_4$.

Before moving on, we first introduce the following lemma.

{\textbf{Lemma 2.}} Given the Laplacian matrix ${L_p} = [{l_{ij}}] \in {{\bf{R}}^{p \times p}}$ , and the diagonal matrix  $\Lambda  = diag({\lambda _1},...,{\lambda _m},0,...,0) \in {{\bf{R}}^{p \times p}}$ where ${\lambda _i} < 0,\forall i = 1,...,m$ , and  ${\lambda _j} = 0,\forall j = m + 1,...,p$ .
Then the matrix
\[A =  - {L_p} + \Lambda \]
is Hurwitz .

Proof.  A system including $A$ can be rewritten as
\[\left\{ \begin{array}{l}
{{\dot x}_1} = {A_1}{x_1} + {A_2}{x_2}\\
{{\dot x}_2} = {A_3}{x_1} + {A_4}{x_2}
\end{array} \right.\]
where
\[\begin{array}{l}
{A_1} =  - \left[ {\begin{array}{*{20}{c}}
{{l_{11}} - {\lambda _1}}&{{l_{12}}}&{...}&{{l_{1m}}}\\
{{l_{21}}}&{{l_{22}} - {\lambda _2}}&{...}&{{l_{2m}}}\\
{...}&{...}&{...}&{...}\\
{{l_{m1}}}&{{l_{m2}}}&{...}&{{l_{mm}} - {\lambda _m}}
\end{array}} \right],\\
{A_2} =  - \left[ {\begin{array}{*{20}{c}}
{{l_{1,m + 1}}}&{...}&{{l_{1,p}}}\\
{...}&{...}&{...}\\
{{l_{m,m + 1}}}&{...}&{{l_{m,p}}}
\end{array}} \right],\\
{A_3} =  - \left[ {\begin{array}{*{20}{c}}
{{l_{m + 1,1}}}&{{l_{m + 1,2}}}&{...}&{{l_{m + 1,m}}}\\
{...}&{...}&{...}&{...}\\
{{l_{p,1}}}&{{l_{p2}}}&{...}&{{l_{pm}}}
\end{array}} \right],\\
{A_4} =  - \left[ {\begin{array}{*{20}{c}}
{{l_{m + 1,m + 1}}}&{...}&{{l_{m + 1,p}}}\\
{...}&{...}&{...}\\
{{l_{p,m + 1}}}&{...}&{{l_{p,p}}}
\end{array}} \right]
\end{array}\].
 In the frequency domain,the above equation can be written as
\[\begin{array}{l}
(s{I_{m \times m}} - {A_1}){x_1}(s) = {x_{10}} + {A_2}{x_2}(s)\\
(s{I_{(p - m) \times (p - m)}} - {A_4}){x_2}(s) = {x_{20}} + {A_3}{x_1}(s)
\end{array}\]
where $x_{10}$ and $x_{20}$  is the initial value.
 We have
 \begin{align}
{x_1}(s) &= {((s{I_{m \times m}} - {A_1}) - {A_2}{(s{I_{(p - m) \times (p - m)}} - {A_4})^{ - 1}}{A_3})^{ - 1}} \nonumber \\
 &\times ({x_{10}} + {A_2}{(s{I_{(p - m) \times (p - m)}} - {A_4})^{ - 1}}{x_{20}})\nonumber
\end{align}
Then the final value of $x_1(t)$ is
\begin{align}
{x_1}(\infty ) &= \mathop {\lim }\limits_{s \to 0} s(((s{I_{m \times m}} - {A_1}) - {A_2}{(s{I_{(p - m) \times (p - m)}} - {A_4})^{ - 1}}\nonumber\\
&{A_3}{)^{ - 1}}({x_{10}} + {A_2}{(s{I_{(p - m) \times (p - m)}} - {A_4})^{ - 1}}{x_{20}}))
\end{align}

Due to
\[{((s{I_{m \times m}} - {A_1}) - {A_2}{(s{I_{(p - m) \times (p - m)}} - {A_4})^{ - 1}}{A_3})^{ - 1}} = \frac{{\bar A{{(s)}^*}}}{{\det (\bar A(s))}}\]
where $\bar A(s) = (s{I_{m \times m}} - {A_1}) - {A_2}{(s{I_{(p - m) \times (p - m)}} - {A_4})^{ - 1}}{A_3}$ , $\det (\bar A(s))$ denotes the determinant of $\bar A(s)$ ,$\bar A{(s)^*}$ denotes the adjoin matrix of $\bar A(s)$ .
It follows from above equation that when $\left| {\bar A(0)} \right| = 0$ ,${x_1}(\infty ) \ne 0$
  , and when $\left| {\bar A(0)} \right| \ne 0$ , ${x_1}(\infty ) = 0$.
Since $\det (\bar A(0)) = \det ( - {L_p} + \Lambda )$ , $\det ({L_p}) = 0$,and $\det (\Lambda ) \ne 0$, we have
 \[\left| {\bar A(0)} \right| \ne 0\].
So  ${x_1}(\infty ) = 0$ and ${x_2}(\infty ) = 0$,It implies that  the system is stale
and $A$ is Hurwitz .%\blacksquare

Using above lemma, for the recursive process 2, we have the following result.

{\textbf{Lemma 3. }}If the system $\sum_3$  and $\sum_4$  satisfy  the assumption 2, the interconnected system composed by  $\sum_3$  and $\sum_4$    can achieve consensus.

\emph{Proof. } For $\sum_3$, define the consensus error of  $x_i$ with respect to the leader $z$  as ${e_i} = {x_i} - z$ , and $e = [{\begin{array}{*{20}{c}}
{{e_1}}&{...}&{{e_n}]}
\end{array}^T}$. Transform  $\sum_3$ into the error system as follows
\begin{equation}  \label{10}
\dot e = Ae - B\dot z
\end{equation}
Solving above equation obtains
\[e(t) = {e^{At}}{e_0} - \int_0^t {{e^{A(t - \tau )}}B\dot zd\tau } \]
By lemma 2, A is Hurwitz, and  there exist $\lambda _1,\lambda _2>0$,  such that ${-\lambda _1}I < A < -{\lambda _2}I$ ,
we have
\begin{equation}  \label{11}
\left\{ \begin{array}{l}
{x_i}(t) \le {e^{ - {\lambda _2}t}}{{\bar e}_0} + z - \int_0^t {{e^{ - {\lambda _2}(t - s)}}} \dot zds\\
{x_i}(t) \ge {e^{ - {\lambda _1}t}}{{\bar e}_0} + z - \int_0^t {{e^{ - {\lambda _1}(t - s)}}} \dot zds
\end{array} \right.
\end{equation}
where  ${\bar e_0}$ is a linear combination of $e_0$ .
Substituting (\ref{11}) into $\sum_4$ yields a cascade structure as follows

\begin{equation}  \label{12}
\dot z \le \sum\limits_{q \in {N_z}}^{} {{K_{qz}}({e^{ - {\lambda _2}t}}{{\bar e}_0} - \int_0^t {{e^{^{ - {\lambda _2}(t - s)}}}} \dot zds)}
\end{equation}
\begin{equation}  \label{13}
\dot z \ge \sum\limits_{q \in {N_z}}^{} {{K_{qz}}({e^{ - {\lambda _1}t}}{{\bar e}_0} - \int_0^t {{e^{^{ - {\lambda _1}(t - s)}}}} \dot zds)}
\end{equation}

\[\dot e = Ae - B\dot z \]

Define  $K = \sum\limits_{q \in {N_z}}^{} {{K_{qz}}} $.For (\ref{12}), we have
\begin{equation}  \label{14}
\dot z \le K{e^{ - {\lambda _2}t}}{\bar e_0} - K\int_0^t {{e^{^{ - {\lambda _2}(t - s)}}}} \dot zds
\end{equation}
Using the similar technique of  Bellman-Gronwall lemma in \cite{Ioannou2006} and the comparison principle \cite{Khalil2002}, we have

\begin{align}
\dot z &\le K{e^{ - {\lambda _2}t}}{{\bar e}_0} - {b_2}{{\bar e}_0}{K^2}{e^{ - {\lambda _2}t}}\int_0^t {{e^{ - {\lambda _2}s}}{e^{{\lambda _2}s}}{e^{ - {b_2}K\int_s^t {{e^{{\lambda _2}\tau }}{e^{ - {\lambda _2}\tau }}d\tau } }}ds} \nonumber\\
 &= K{e^{ - {\lambda _2}t}}{{\bar e}_0} - {b_2}{{\bar e}_0}{K^2}{e^{ - {\lambda _2}t}}\int_0^t {{e^{ - {b_2}K(t - s)}}ds} \nonumber\\
 &= K{e^{ - {\lambda _2}t}}{{\bar e}_0} - {b_2}{{\bar e}_0}{K^2}{e^{ - ({\lambda _2} + {b_2}K)t}}\int_0^t {{e^{{b_2}Ks}}ds} \nonumber\\
 &= K{e^{ - {\lambda _2}t}}{{\bar e}_0} - \frac{{{b_2}{{\bar e}_0}{K^2}{e^{ - ({\lambda _2} + {b_2}K)t}}}}{{{b_2}K}}({e^{{b_2}Kt}} - 1)\nonumber\\
 &= K{e^{ - {\lambda _2}t}}{{\bar e}_0} - {b_2}{{\bar e}_0}K({e^{ - {\lambda _2}t}} - {e^{ - ({\lambda _2} + {b_2}K)t}}) \nonumber
\end{align}

Due to ${\lambda _2},{b_2},K > 0$, we have
\[{\beta _2}({e_0},t) = K{e^{ - {\lambda _2}t}}{\bar e_0} - {b_2}{\bar e_0}K({e^{ - {\lambda _2}t}} - {e^{ - ({\lambda _2} + {b_2}K)t}}) \in KL.\]
Similarly, for (\ref{13})  we have  $\dot z \ge {\beta _1}({e_0},t)$
where ${\beta _1}({e_0},t) = K{e^{ - {\lambda _1}t}}{\bar e_0} - {b_1}{\bar e_0}K({e^{ - {\lambda _1}t}} - {e^{ - ({\lambda _1} + {b_1}K)t}}) \in KL.$
Therefore,
\[{\beta _1}({e_0},t) \le \dot z \le {\beta _2}({e_0},t)\]
It is clear that  $\mathop {\lim }\limits_{t \to \infty } \dot z(t) = 0$  and
\[z(0) + \int_0^t {{\beta _1}({e_0},s)ds}  \le z(t) \le z(0) + \int_0^t {{\beta _2}({e_0},s)ds} \]
For equation (\ref{10}), the error system is stable, thus
\[\mathop {\lim }\limits_{t \to \infty } \left\| {{x_i}(t) - z} \right\| = 0\]
That is $z(t) \to c$, $\forall t \to \infty$, where $c$ is a constant determined by the initial values of $\sum_3$ and $\sum_4$. Therefore, the interconnected system  achieves consensus.%\blacksquare

\subsection{Combination of  Cascade Structure and Interconnected Structure}

This subsection will investigate a more complex structure, the combination of the cascade structure and the interconnected structure.

\begin{center}
\includegraphics [scale=0.3]{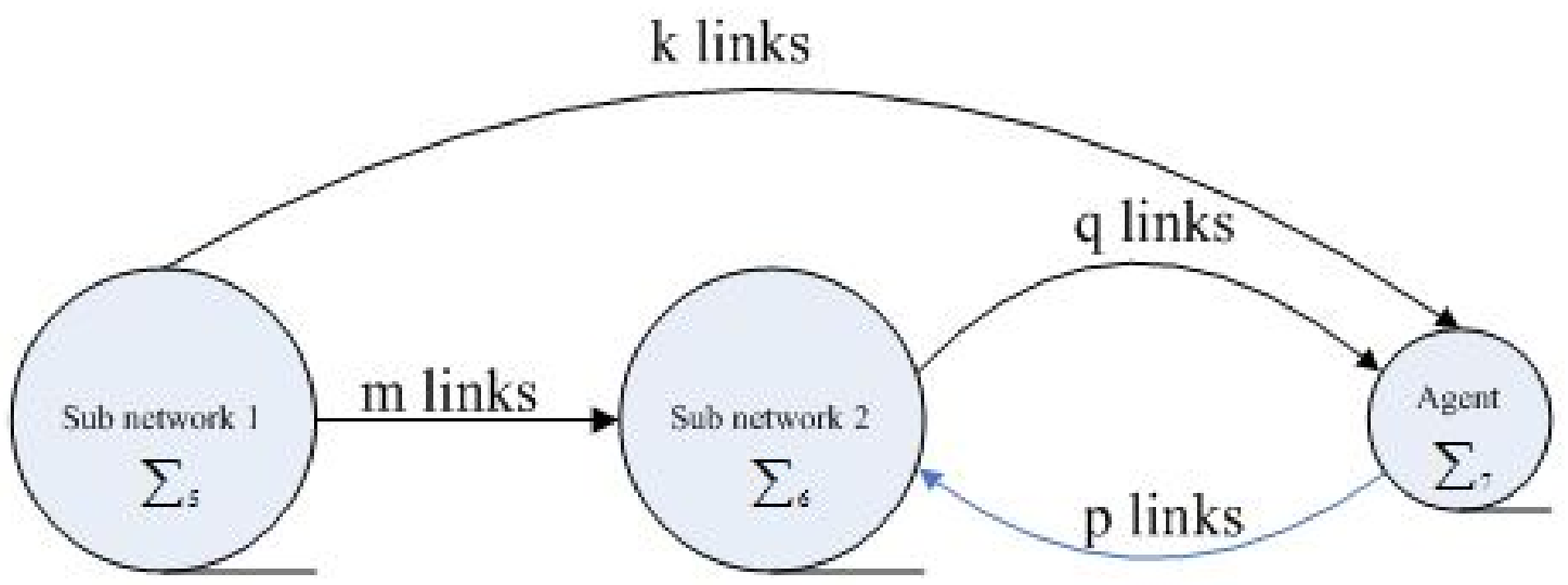}
\\
{\fontsize{7.3pt}{11.6pt}\selectfont
Fig.~8~~The blended structure of multi-agent systems.}
\end{center}

Consider the interconnected system in the above subsection, if $\sum_4$ is not the root of any spanning tree of the interconnected system, then the topology can be illustrate by the Fig.8, where  sub-network 2(denoted as $\sum_6$  )  and the agent (denoted as $\sum_7$  ) make up an interconnected structure, and sub-network 1(denoted as $\sum_5$ )  and  the interconnected structure  make up a cascade structure. All of them are modeled as follows.

$\sum_5$:
\[{\dot x_i} = \sum\limits_{v \in {N_i}} {K_{vi}^{}({x_v} - {x_i})} ,\forall i = 1,...,{n_1}\]
$\sum_6$:
\[{\dot x_j} = \sum\limits_{{w_1} \in {N_{j1}}} {K_{{w_1}j}^{}(x_{{w_1}}^{} - {x_j})}  + \sum\limits_{{w_2} \in {N_{j2}}} {K_{{w_2}j}^{}(x_{{w_2}}^{} - {x_j})}  + K_{zj}^{}(z - {x_j})\]
\[\forall j =  1,...,p, x_{{w_2}}\in \sum_5 \]
\[{\dot x_k} = \sum\limits_{{p_1} \in {N_{k1}}} {K_{{p_1}k}^{}({x_{{p_1}}} - {x_k})}  + \sum\limits_{{p_2} \in {N_{2k}}} {K_{{p_2}k}^{}({x_{{p_2}}} - {x_k})} \]
\[\forall k = p + 1,...,{n_2},x_{{p_2}}\in \sum_5\]
$\sum_7$:
\[\dot z = \sum\limits_{{q_1} \in {N_{z1}}}^{} {{K_{{q_1}}}({x_{{q_1}}} - z)}  + \sum\limits_{{q_2} \in {N_{z2}}}^{} {{K_{{q_2}}}({x_{{q_2}}} - z)} \]
\[\forall {q_1} = 1,...,q,x_{{q_1}}\in \sum_6,x_{{q_2}}\in \sum_5\]
where at least one of $K_{{w_2}j}$ , $K_{{p_2}j}$ is not  zero, which means there exists at least one link from $\sum_5$ to $\sum_6$ .

\emph{\textbf{Assumption 3.}} System  $\sum_5$   achieves consensus, and  $\sum_6$ after removing the links from $\sum_5$ and $\sum_7$ can achieves consensus. The interconnected structure of  $\sum_6$ and $\sum_7$  satisfies assumption 2.

We have the following result.

{\textbf{Lemma 4.}} If systems   $\sum_5$ , $\sum_6$,and $\sum_7$   satisfy  the assumption 3, the multi-agent system composed by  $\sum_5$ , $\sum_6$,and $\sum_7$   can achieve consensus.

\emph{Proof.} Because  $\sum_5$ makes up a cascade structure with   $\sum_6$ and $\sum_7$  ,  we can first analyze the system composed by   $\sum_6$ and $\sum_7$  . Define $x_v = [x_i]^{n\times 1},\forall x_i \in \sum_5$  Then
the solution of the interconnected system ( $\sum_6$ and $\sum_7$ ) is
\begin{equation} \label{15}
x(t) = {e^{At}}{x_0} - \int_0^t {{e^{A(t - \tau )}}B{x_{v}}(s)d\tau }
\end{equation}
Since $\sum_5$ can achieve consensus, there exists ${\beta _p} \in KL$ such that
\[ - {\beta _p}({x_0},t) + {x_e} \le {x_{v}}(t) \le {\beta _p}({x_0},t) + {x_e}\]
Substituting above equation into (\ref{15}) yields

\begin{align}
x(t)& = {e^{At}}{x_0} - \int_0^t {{e^{A(t - \tau )}}B({\beta _p}({x_0},s) + {x_e})d\tau } \nonumber \\
 &= {e^{At}}{x_0} - \int_0^t {{e^{A(t - \tau )}}B{\beta _p}({x_0},s)d\tau }  - {x_e}\int_0^t {{e^{A(t - \tau )}}Bd\tau }\nonumber\\
 &= {e^{At}}{x_0} - \int_0^t {{e^{A(t - \tau )}}B{\beta _p}({x_0},s)d\tau } {\rm{ + }}{x_e}{A^{ - 1}}\int_0^t {d{e^{A(t - \tau )}}d\tau } B\nonumber\\
 &= {e^{At}}{x_0} - \int_0^t {{e^{A(t - \tau )}}B{\beta _p}({x_0},s)d\tau }  - {x_e}{A^{ - 1}}{e^{At}}B + {x_e}{A^{ - 1}}B\nonumber
\end{align}

It is apparent that the interconnected system ($\sum_6$ and $\sum_7$) is ISS w.r.t. $\sum_5$, and the gain is 1,so that ${A^{ - 1}}B = {1_n}$, and ${e^{At}} \in KL$.

Then the system is consensus.%\blacksquare

The three topology multi-agent systems are basic elements to construct the complex system, the next section will show that any topology multi-agent system with a spanning tree can achieve consensus.

\section{Consensus Analysis}

In this section, we will combine the results of section3 and 4 to prove any topology multi-agent system can achieve consensus.  Then we have the following main result.

{\textbf{Theorem 1. }} The protocol algorithm (2) can guarantee the dynamical system (1) achieves consensus, if the interconnection graph has a spanning tree.

\emph{{Proof.}} The proof completely follows the Procedure 1.Suppose the graph of the system is $g$.
Since there exists at least one spanning tree in it, and then arbitrarily choose one ,and
denote it as $g_s$ . Set $E_{\bar s}^0 = E - {E_s}$ , ${{\bf{G}}^0} = \emptyset $ , and ${\bf{G}}_s^0 = \emptyset $ .

\begin{description}
  \item[(1)] Initialize graph $G_{}^1$
  \end{description}

Choose the root agent  $x_1$  of $g_s$ and adding it to $G^0$ and $G_{s}^0$  yields $G_{}^1$  and $G_{s}^1$ ,modeled as
\begin{equation} \label{15}
\dot x_1 = 0
\end{equation}

\begin{description}
  \item[(2)] Add agent $x_2$ to $G_{}^1$  and $G_{s}^1$  so as to obtain  $G_{}^2$  and $G_{s}^2$.
  \end{description}
  By  the interconnection configuration, there are two cases.
\begin{itemize}
  \item Case 1:  there is only  one edge from  $G_{}^1$ to $x_2$ .
\end{itemize}

Adding an order edge from  $G_{}^1$ to $x_2$ implies adding an ISP to $x_2$. The new graph $G_{}^2$ is a cascade structure as follows
\[\left\{ \begin{array}{l}
{{\dot x}_1} = 0\\
{{\dot x}_2} = {K_{12}}({x_1} - {x_2})
\end{array} \right.\]
By Lemma 1,  $x_1$ is a constant determined by its initial value, therefore,  $G_{}^2$ can achieve consensus.
\begin{itemize}
  \item Case 2:  there exists the  bidirected edge from   $G_{}^1$ to $x_2$ .
\end{itemize}
On the basis of case 1, add  another edge from  $x_1$ to $G_{}^1$   , then  we have
\[\left\{ \begin{array}{l}
{{\dot x}_1} = {K_{21}}({x_2} - {x_1})\\
{{\dot x}_2} = {K_{12}}({x_1} - {x_2})
\end{array} \right.\]
By lemma 3, the interconnected system $G^2$ can achieve consensus .

\begin{description}
  \item[(3)]Choose the $k$-th agent  $z$ in $g_s$  according to the width-first sequence, and  add it to   $G_{}^{k-1}$ and $G_{s}^{k-1}$  ,where $k = 2,...,n$, so as to obtain $G_{}^{k}$ and $G_{s}^{k}$.
  \end{description}
By the specific topology, there are three cases:
\begin{itemize}
  \item Case 1:  There exists m order edges from   $G_{}^{k-1}$ to $z$ ,so that the new graph $G^k$ has a cascade structure  like Fig.6, modeled as  $\sum_1$ and $\sum_2$. By lemma 1, $G^k$ can achieve consensus.
\end{itemize}

\begin{itemize}
  \item Case 2:  The interconnection between $G^{k-1}$  and $z$  belongs to the
interconnected structure like Fig.7, and satisfies assumption 2 , which can be modeled as $\sum_3$  and $\sum_4$ .Hence , by lemma 3, $G^k$ can achieve consensus.
\end{itemize}

\begin{itemize}
  \item Case 3:  The interconnection between   $G^{k-1}$  and $z$   belongs to
a blended structure like Fig.8, and satisfies assumption 3, which can be modeled as $\sum_5$  ,$\sum_6$  and $\sum_7$ .Hence , by lemma 4,  $G^k$ can achieve consensus.
\end{itemize}

 In conclusion,  $G^k$  can achieve consensus.

(4)	At last, repeat  (3)  till ${E_{\bar s}} = \emptyset $ , then by Proposition 1,
we reconstruct a new graph ${{\bf{G}}^n} = ({V^n},{E^n})$  and $G^n$  is the original graph $g$ . Since  the protocol algorithm (2) can guarantee $G^k$  achieve consensus at every step, so is $G^n$  .
%\blacksquare

\section{Conclusion}

In this paper, based on the structural perspective, we propose a new framework for the consensus analysis of linear multi-agent systems.  Since the diversity of the topology of graph and consider all details of every agent,
the traditional approach tends to be confined by the specific topology. Otherwise, the approach in this paper focuses on every agent and takes full advantage of its ISS property so that it can be used for any topology multi-agent system. Besides, this paper provides a structural explain  how the multi-agent system achieves consensus, that is any topology multi-agent system can be transformed into a cascade system whose  first agent converges to a constant determined by initial values of some agents. Further more, this framework can be easily generalized to other  systems, like the high-order line or the nonlinear system, the discrete-time system, and the time-delay system, etc. In the future, we will make attempt to generalize it to the multi-agent with nonlinear protocols.

\end{document}